\documentclass[twocolumn,showpacs,preprintnumbers,amsmath,amssymb,floatfix]{revtex4}
\usepackage{graphicx}
\usepackage{dcolumn}
\usepackage[latin1]{inputenc}
\usepackage{xcolor}
\usepackage{epstopdf}

\begin{document}

\title{The semiconducting and metallic phases of conjugated polymers} 
\author{Miguel Lagos}
\email{mlagos@utalca.cl}
\affiliation{Facultad de Ingenier{\'\i}a, Universidad de Talca,
Campus Los Niches, Camino a los Niches km 1, Curic{\'o}, Chile}

\date{February 22, 2020}

\begin{abstract}
Recently a variety of $\pi$--conjugated polymers have been
developed and essayed for a number of applications such as organic
light--emitting diodes, organic field-effect transistors, organic
photovoltaics, and sensors. Is central for these applications the
semiconductor character of the pure materials, which can turn into
metallic conductivity by local oxydation or reduction. Many very recent
experiments show two characteristic peaks in the UV--Vis excitation
spectra of the conjugated polymers of interest, both lying in the gap
between the energies of the bonding $\pi$ and antibonding $\pi^*$ bands
and having excitation energies in a ratio ranging from 1.4 to 1.7. The
issue is that $\pi$--electrons are paired in covalent orbitals which
interact strongly between them and with the ionic cores, thus being far
from the extended quasi--free independent one--electron states assumed
by the theory of inorganic semiconductors. A model yielding a mechanism
for many--body conduction of charge and semiconducting properties of
the undoped material is introduced here. The model yields two new flat
bands of excited bonding states of  Bose--Einstein statistics able of
charge transport. The two bands explain well the characteristic pair of
maxima of the UV--Vis excitation spectra of the conductive conjugated
polymers.   
\end{abstract}

{\pacs{no longer in use}

\maketitle

In opposition to the importance of their potential or actual
applications \cite{LeKimYoon}, the underlying science that give rise to
the semiconducting and conducting properties of conjugated polymers is
yet in an early stage of understanding. In comparison with inorganic
semiconductors, relatively little is known about the physical origin
of the electronic properties of the conjugated polymers, and even the
precise nature of the semiconductor excitations remains yet uncertain.
Inorganic semiconductors are characterized by the long range spatial
order of a single crystal, where delocalized non--interacting
quasi--free electrons evolve in the periodic field of the crystalline
lattice. On the contrary, in organic semiconductors the strict spatial
order is often reduced to one dimension because the different polymeric
chains are rarely in ordered arrays. More importantly, the electrons
are paired in covalent highly localized $\sigma$-- or $\pi$--orbitals
which interact strongly between them and with the ionic cores, and are
then far from being in delocalized free or quasi--free states.

Conjugated polymers are polymers constituted by chains of identical
molecular structures held together by a sequence of bonds that
alternates single and double covalent bonds. The double bonds combine
a $\sigma$-- and a $\pi$--bond, whereas the single ones are
$\sigma$--bonds. Hence the backbone is essentially the more stable
uniform sequence of $\sigma$--bonds. The simplest conjugate polymer is
polyacetylene, for which the periodic molecular structure is just a
carbon atom with one of its valence electrons in covalent bond with an
hydrogen atom. The metallic state is reached by doping through local
oxidation or reduction. Thin films of conjugated polymers are
attracting considerable interest because of their varied actual and
potential applications in electronic and optoelectronic devices, such
as transistors, photodiodes, organic photovoltaic (OPV) devices,
organic light-emitting diodes (OLEDs) and many others, combined with
their versatility and ease low cost fabrication. Even the use of paper
as a substrate for organic transistors is now being investigated
\cite{Zschieschang}. Devices combining the injected electrons from one
electrode and holes from the other electrode of a two--layer organic
emitting diode were found to have a good electroluminescent efficiency
with appropriately low operation voltages
\cite{Tang,Adachi,Burroughes,Braun}. Now the use of organic materials
as the active semiconductors in electronic flat panel displays are in
large scale production and commercialization. Conjugated polymers are
intrinsically stable upon excitation by an applied voltage or photon
capture in either light emission or harvesting devices. The alternating
$\pi$--bonds participate actively in the electronic processes leaving
intact the primary structure, constituted by the uniform backbone of
$\sigma$--bonds. This backbone provides the necessary stability against
degradation by the energy transfers demanded by the operation of the
devices.

In general, carbon atoms of the backbone take electrons from
neighboring atoms to constitute pairs with their four valence
electrons in order to form particularly stable octets. Hence the
spatial electronic pairing inherent to the $\pi$--bonds is a primary
condition of the conjugate polymers. Models for the charge transport
along the polymer chain must take this into account, and presume that
the occurrence of any one--electron elementary process demands a too
large activation energy. Bearing this in mind, it is proposed here an
essentially many--body mechanism of charge transport which meets the
conditions posed by the structure of the conjugated polymers much
better than the conventional scheme developed for inorganic conductors
and semiconductors. The results agree well with direct observations of
the electronic structure of polymer conductors and semiconductors by
the absorption of UV--Vis photons \cite{Chen,Yao,Mangalore,Li,Lee,Park}.

The model put forward here is consistent with the pioneer ideas of Su,
Schrieffer and Heeger \cite{Su1,Su2,Heeger}, but with explicit
recognition of the importance of the electrostatic repulsion between
$\pi$--orbitals in neighbouring sites \cite{Zhang}. Chain dimerization
is taken as granted and included as an implicit condition. Charge
carriers in the conducting states have the same structure than the
solitons of Su, Schrieffer and Heeger, but with some subtle differences.
The model Hamiltonian for the polymer chain is written as 

\begin{equation}
H=H_0+H_{\text{C}}+H_{\text{T}},
\label{E1}
\end{equation}

\noindent
where

\begin{equation}
H_0=\sum_l\epsilon\,\big( n_{l\uparrow}+n_{l\downarrow}\big),
\label{E2}
\end{equation}

\noindent
$\epsilon$ denotes the energy per electron of the two--electron
$\pi$--orbital and $n_{l\uparrow}=c_{l\uparrow}^\dagger c_{1\uparrow}$,
$n_{l\downarrow}=c_{l\downarrow}^\dagger c_{l\downarrow}$, are the
occupation operators of the two one--electron states at sites
$l=-N/2,-N/2+1,\dots ,N/2$. The term $H_C$ stems from the Coulomb
repulsion between $\pi$--orbitals located in adjacent sites of the
polymer chain. Zhang et al.~has proven that the $\pi$--$\pi$
electrostatic repulsion has significant effect in conjugated systems,
and the $\pi$--$\pi$ Pauli repulsion plays a secondary but
nonnegligible role \cite{Zhang}. Hence

\begin{equation}
H_{\text{C}}=
\sum_l U\big(n_{l+1\uparrow}+n_{l+1\downarrow}\big)
\big(n_{l\uparrow}+n_{l\downarrow}\big).
\label{E3}
\end{equation}

\noindent
The two--electron tunneling term

\begin{equation}
H_{\text{T}}=
\sum_l V\big(c_{l+1\uparrow}^\dagger c_{l+1\downarrow}^\dagger
c_{l\downarrow}c_{l\uparrow}
+c_{l\uparrow}^\dagger c_{l\downarrow}^\dagger
c_{l+1\downarrow}c_{l+1\uparrow}\big)
\label{E4}
\end{equation}

\noindent
accounts for quantum fluctuations of the $\pi$--bonds between
neighbouring sites. The underlying principle is that the low energy
$N$--electron states of the chain can be expressed as combinations of
two--electron $\pi$--states localized every other site along the chain.
Hence any term of the Hamiltonian not having this general structure
will give vanishing contribution when operating on the paired states. 

The transformed new dynamical variables

\begin{equation}
\begin{aligned}
& s_1(l)=\frac12\big( c_{l\uparrow}^\dagger c_{l\downarrow}^\dagger
+c_{l\downarrow}c_{l\uparrow}\big)\\
& s_2(l)=\frac1{2i}\big( c_{l\uparrow}^\dagger c_{l\downarrow}^\dagger
-c_{l\downarrow}c_{l\uparrow}\big)\\
& s_3(l)=\frac12\big( n_{l\uparrow}+n_{l\downarrow}-1\big)
\end{aligned}
\label{E5}
\end{equation}

\noindent
satisfy commutation relations of the components of an angular momentum

\begin{equation}
[s_1,s_2]=is_3 \quad [s_2,s_3]=is_1 \quad [s_3,s_1]=is_2.
\label{E6}
\end{equation}

\noindent
As the number operators $c_l^\dagger c_l$ have eigenvalues 0 and 1,
the eigenvalues of $s_3$ are $-1/2$, $0$ and $1/2$. The eigenvalue $0$
conveys the breaking of a covalent bond, which has a large energy
cost. Taking it as infinite the operators $s_1$, $s_2$, and $s_3$
behave as the components of a spin 1/2. The Hamiltonian $H$ then takes
the general form of the Hamiltonian of an anisotropic spin 1/2
antiferromagnetic Heisenberg model

\begin{equation}
\begin{aligned}
H=\,&\big(\epsilon+2U\big)\sum_l\big[ 2s_3(l)+1\big]
+4U\sum_l\bigg( s_3(l+1)s_3(l)\\
&+\frac{V}{2U}\big[ s_1(l+1)s_1(l)+s_2(l+1)s_2(l)\big]\bigg)
\label{E7}
\end{aligned}
\end{equation}

\noindent
which in terms of the ladder operators $s_+=s_1+is_2$ and
$s_-=s_1-is_2$ can be rewritten as

\begin{equation}
\begin{aligned}
H=\,&\big(\epsilon+2U\big)\sum_l\big[ 2s_3(l)+1\big]
+4U\sum_l\bigg( s_3(l+1)s_3(l)\\
&+\frac{V}{4U}\big[ s_+(l+1)s_-(l)+s_+(l)s_-(l+1)\big]\bigg).
\label{E8}
\end{aligned}
\end{equation}

\noindent
As just full occupation will be considered, the first sum in the right
hand side of this equation is a constant and can be disregarded.

The use of fermion or angular moment operators are two formally
equivalent alternatives to deal with the model put forward here.
Translating them into the fermion scheme, the published techniques for
dealing with the anisotropic Heisenberg antiferromagnetic chain \cite{LagosCabrera,LagosKiwiGaglianoCabrera,CabreraLagosKiwi},
can be applied here. In this spirit define first the operators

\begin{equation}
\phi_{\text{e}}^\dagger =\sqrt{\frac{2}{N}}
\sum_{\text{even }l} c_{l+1\uparrow}^\dagger c_{l+1\downarrow}^\dagger
c_{l\downarrow}c_{l\uparrow}+
\frac{\alpha}{2}\sqrt{\frac{N}{2}},
\label{E9}
\end{equation}

\begin{equation}
\phi_{\text{o}}^\dagger =\sqrt{\frac{2}{N}}
\sum_{\text{odd }l} c_{l\uparrow}^\dagger c_{l\downarrow}^\dagger
c_{l+1\downarrow}c_{l+1\uparrow}+
\frac{\alpha}{2}\sqrt{\frac{N}{2}},
\label{E10}
\end{equation}

\noindent
where $\alpha =V/(2U)$. Recalling the elementary identities $[A,BC]=
[A,B]C+B[A,C]$ and $[A,BC]=\{A,B\}C-B\{A,C\}$ between commutators and
anticommutators in standard notation, one can readily show they have
the commutation properties

\begin{equation}
\begin{aligned}
&[\phi_{\text{e}},\,\phi_{\text{e}}^\dagger ]
=\frac{2}{N}\sum_{\text{even }l}
\big[ n_{l\uparrow} n_{l\downarrow}
\big( 1-n_{l+1\uparrow}-n_{l+1\downarrow} \big)\\
&\phantom{ababab}-n_{l+1\uparrow} n_{l+1\downarrow}
\big( 1-n_{l\uparrow}-n_{l\downarrow} \big)\big],
\label{E11}
\end{aligned}
\end{equation}

\begin{equation}
\begin{aligned}
&[\phi_{\text{o}},\,\phi_{\text{o}}^\dagger ]
=-\frac{2}{N}\sum_{\text{odd }l}
\big[ n_{l\uparrow} n_{l\downarrow}
\big( 1-n_{l+1\uparrow}-n_{l+1\downarrow} \big)\\
&\phantom{ababab}-n_{l+1\uparrow} n_{l+1\downarrow}
\big( 1-n_{l\uparrow}-n_{l\downarrow} \big)\big],
\label{E12}
\end{aligned}
\end{equation}

\begin{equation}
[\phi_{\text{e}},\,\phi_{\text{o}}]\equiv 0,
\label{E13}
\end{equation}

\begin{equation}
\begin{aligned}
&[H_C,\,\phi_{\text{e}}^\dagger]=
2U\sqrt{\frac{2}{N}}\sum_{l\text{ even}}
c_{l+1\uparrow}^\dagger c_{l+1\downarrow}^\dagger
c_{l\downarrow}c_{l\uparrow}\\
&\phantom{abababab}\times (n_{l+2\uparrow}+n_{l+2\downarrow}
-n_{l+1\uparrow}-n_{l+1\downarrow}\\
&\phantom{abababab}+n_{l\uparrow}+n_{l\downarrow}-n_{l-1\uparrow}
-n_{l-1\downarrow}-2),
\label{E14}
\end{aligned}
\end{equation}

\noindent
and

\begin{equation}
\begin{aligned}
&[H_C,\,\phi_{\text{o}}^\dagger]=
2U\sqrt{\frac{2}{N}}\sum_{l\text{ odd}}
c_{l\uparrow}^\dagger c_{l\downarrow}^\dagger
c_{l+1\downarrow}c_{l+1\uparrow}\\
&\phantom{abababab}\times (-n_{l+2\uparrow}-n_{l+2\downarrow}
+n_{l+1\uparrow}+n_{l+1\downarrow}\\
&\phantom{abababab}-n_{l\uparrow}-n_{l\downarrow}+n_{l-1\uparrow}
+n_{l-1\downarrow}-2).
\label{E15}
\end{aligned}
\end{equation}

\noindent
The commutators (\ref{E10}) and (\ref{E11}) go to Bose commutation
relations

\begin{equation}
[\phi_{\text{e}},\,\phi_{\text{e}}^\dagger ]
=[\phi_{\text{o}},\,\phi_{\text{o}}^\dagger ]=1
\label{E16}
\end{equation}

\noindent
in the asymptotic limit of high conjugation

\begin{equation}
n_{l\uparrow}=n_{l\downarrow}\rightarrow
\begin{cases}
1,\text{ if }l\text{ even}\\
0,\text{ if }l\text{ odd.}
\end{cases}
\label{E17}
\end{equation}

\noindent
Also, in the same limit,

\begin{equation}
[H_C,\phi_{\text{e,o}}^\dagger ]=
4U\bigg(\phi_{\text{e,o}}^\dagger
-\frac{\alpha}{2}\sqrt{\frac{N}{2}}\bigg).
\label{E18}
\end{equation}

\noindent
Noticing that

\begin{equation} 
H_T=\sqrt{\frac{N}{2}}\, V(\phi_{\text{e}}^\dagger
+\phi_{\text{o}}^\dagger +\phi_{\text{e}}+\phi_{\text{o}})
-N\alpha V,
\label{E19}
\end{equation}

\noindent
disregarding $H_0$ one has from
Eqs.~(\ref{E1}), (\ref{E18}) and (\ref{E19}) that

\begin{equation}
[H,\phi_{\text{e}}^\dagger ]=4U\phi_{\text{e}}^\dagger ,\qquad
[H,\phi_{\text{o}}^\dagger ]=4U\phi_{\text{o}}^\dagger .
\label{E20}
\end{equation}

\noindent
Hence in the limit of strong conjugation (\ref{E17}) the operators
$\phi_{\text{e}}$ and $\phi_{\text{o}}$ are ladder operators.

The ground state $|g\rangle$ of $H$ must satisfy

\begin{equation}
\phi_{\text{e}}|g\rangle =\phi_{\text{o}}|g\rangle =0.
\label{E21}
\end{equation}

Defining now $|\mathcal{N}\rangle$ as the chain of bare
$\pi$--electronic states (the N{\'e}el state in the spin
representation) 

\begin{equation}
|\mathcal{N}\rangle =\prod_{\text{even }l}
c_{l\uparrow}^\dagger c_{l\downarrow}^\dagger |0\rangle
\label{E22}
\end{equation}

\noindent
where $|0\rangle$ is the vacuum, and

\begin{equation}
\begin{aligned}
\Lambda &=\sqrt{\frac{N}{2}}\big(\phi_{\text{e}}^\dagger
+\phi_{\text{o}}^\dagger -\phi_{\text{e}}-\phi_{\text{o}}\big)\\
&=\sum_l (-1)^l \big(
c_{l+1\uparrow}^\dagger c_{l+1\downarrow}^\dagger
c_{l\downarrow}c_{l\uparrow}
-c_{l\uparrow}^\dagger c_{l\downarrow}^\dagger
c_{l+1\downarrow}c_{l+1\uparrow} \big),
\label{E23}
\end{aligned}
\end{equation}

\noindent
it can be shown that the ground state of $H$ in the asymptotic limit
(\ref{E17}) is

\begin{equation}
|g\rangle
=\exp\bigg(-\frac{\alpha}{2}\Lambda\bigg)|\mathcal{N}\rangle .
\label{E24}
\end{equation}

\noindent
To demonstrate this, notice that it can be proven by complete induction
that in the limit (\ref{E17}) one has that

\begin{equation}
[\phi_{\text{e,o}},
(\phi_{\text{e}}^\dagger +\phi_{\text{o}}^\dagger -\phi_{\text{e}}
-\phi_{\text{o}})^n]
=n(\phi_{\text{e}}^\dagger +\phi_{\text{o}}^\dagger -\phi_{\text{e}}
-\phi_{\text{o}})^{n-1},
\label{E25}
\end{equation}

\noindent
hence for any analytic function $F$ with derivative $F^\prime $ 

\begin{equation}
[\phi_{\text{e,o}},F(\phi_{\text{e}}^\dagger
+\phi_{\text{o}}^\dagger -\phi_{\text{e}}
-\phi_{\text{o}})]=F^\prime (\phi_{\text{e}}^\dagger
+\phi_{\text{o}}^\dagger -\phi_{\text{e}}
-\phi_{\text{o}}).
\label{E26}
\end{equation}

\noindent
Applying this property with $F$ substituted by the exponential function
appearing in Eq.~(\ref{E24}) and the definitions (\ref{E9}) and
(\ref{E10}), it can be readily shown that $|g\rangle$ satisfies
Eqs.~(\ref{E21}) and is then the ground state in the limiting case
(\ref{E17}). Notice that the ground state (\ref{E24}) is not
perturbative because of the large factor $\sqrt{N/2}$ multiplying the
sum (\ref{E23}) defining $\Lambda$.

To determine the ground state energy consider the commutation property
of $\Lambda$ 

\begin{equation}
\begin{aligned}
&[c_{l\uparrow}^\dagger c_{l\downarrow}^\dagger ,\Lambda]\\
&=(-1)^l\big( c_{l+1\uparrow}^\dagger c_{l+1\downarrow}^\dagger
+c_{l-1\uparrow}^\dagger c_{l-1\downarrow}^\dagger\big)
\big( n_{l\uparrow}+n_{l\downarrow}-1\big),
\end{aligned}
\label{E27}
\end{equation}

\noindent
which after iterating $\nu$ times in the asymptotic limit (\ref{E17})
reads

\begin{equation}
\begin{aligned}
&[[\dots [c_{l\uparrow}^\dagger c_{l\downarrow}^\dagger ,\Lambda],
\Lambda ],\dots ,\Lambda]_{\nu\text{ times}}\\
&\phantom{abab}
=(-1)^{l\nu}(-1)^{\nu (\nu +1)/2}\big(\tau +\tau^{-1}\big)^{\nu}
c_{l\uparrow}^\dagger c_{l\downarrow}^\dagger ,
\end{aligned}
\label{E28}
\end{equation}

\noindent
where $\tau$ is the translation operator to the next site:

\begin{equation}
\tau c_{ls}=c_{l+1s}\quad\tau^{-1} c_{ls}=c_{l-1s}\quad
s=\uparrow ,\downarrow .
\label{E29}
\end{equation}

\noindent
Combining this with the identity

\begin{equation}
\begin{aligned}
e^{-B}A\, e^B \equiv &A+\frac{1}{1!}[A,B]+\frac{1}{2!}[[A,B],B]\\
&+\frac{1}{3!}[[[A,B],B],B]+\cdots ,
\label{E30}
\end{aligned}
\end{equation}

\noindent
the generating function of the modified Bessel functions $I_\nu (z)$

\begin{equation}
\begin{aligned}
\exp\bigg[\frac{z}{2}\big(\tau +\tau^{-1}\big)\bigg]
=\sum_{\nu =-\infty}^\infty I_\nu (z)\tau^\nu ,
\label{E31}
\end{aligned}
\end{equation}

\noindent
and the property $J_\nu (z)=i^{-\nu}I_\nu(iz)$, where $J_\nu$ is the
unmodified Bessel function, it can be shown that

\begin{equation}
\begin{aligned}
&\exp\bigg(\frac{\alpha}{2}\Lambda\bigg)
c_{l\uparrow}^\dagger c_{l\downarrow}^\dagger
\exp\bigg(-\frac{\alpha}{2}\Lambda\bigg)\\
&=\sum_{\nu =-\infty}^\infty (-1)^{l\nu} (-1)^{\nu (\nu +1)/2}
J_\nu (\alpha )\, c_{l+\nu\uparrow}^\dagger
c_{l+\nu\downarrow}^\dagger .
\label{E32}
\end{aligned}
\end{equation}

\noindent
Eq.~(\ref{E32}) together with Eq.~(\ref{E24}) are useful to calculate
expectation values, as the mean occupation of a $\pi$--orbital 

\begin{equation}
\langle g|(n_{l\uparrow}+n_{l\downarrow})|g\rangle
= 1+(-1)^l J_0(2\alpha ),
\label{E33}
\end{equation}

\noindent
or the short range correlation coefficient

\begin{equation}
\begin{aligned}
\frac{1}{N}\langle g|
\sum_l\big( & n_{l+1\uparrow}+n_{l+1\downarrow}\big)
\big( n_{l\uparrow}+n_{l\downarrow}\big)
|g\rangle \\
& = 1-[J_0(2\alpha )]^2,
\label{E34}
\end{aligned}
\end{equation}

\noindent
or the energy of the ground state $E_g = \langle g|\big( H_C+H_T\big)
|g\rangle $

\begin{equation}
E_g =NU\big[1-\big( J_0(2\alpha )\big)^2 +2\alpha J_1(2\alpha )\big].
\label{E35}
\end{equation}

\noindent
In obtaining Eqs.~(\ref{E33}), (\ref{E34}) and (\ref{E35}) use
was made of Neumann's addition formulas of the Bessel functions
and Graf's generalization of them \cite{Watson}. The previous results
expressed in terms of the Bessel functions look elegant, but care must
be taken in that they hold in the asymptotic limit (\ref{E17}), which
occurs for small enough $\alpha$. Up to the second order in $\alpha$
one has that

\begin{equation}
E_g =4NU(\alpha^2 +0(\alpha^4)).
\label{E36}
\end{equation}

By the properties (\ref{E20}) of the ladder operators
$\phi_{\text{e}}^\dagger$ and $\phi_{\text{o}}^\dagger$ the set of
vectors

\begin{equation}
|n_{\text{e}}\, n_{\text{o}}\rangle
=\dfrac{(\phi_{\text{e}}^\dagger)^{n_{\text{e}}}}{\sqrt{n_{\text{e}}!}}
\dfrac{(\phi_{\text{o}}^\dagger)^{n_{\text{o}}}}{\sqrt{n_{\text{o}}!}}
\,|g\rangle , \quad
n_{\text{e}},\, n_{\text{o}}=0,1,2,3,\dots
\label{E37}
\end{equation}

\noindent
are a set of eigenvectors of $H$, with eigenenergies

\begin{equation}
E_{n_{\text{e}}n_{\text{o}}}=4(n_{\text{e}}+n_{\text{o}})U+E_g.
\label{E38}
\end{equation}

The theoretical framework described up to this point would not be
complete without observing that the ground state is twofold degenerate.
In effect, the state

\begin{equation}
|\bar{g}\rangle
=\exp\bigg(\frac{\alpha}{2}\Lambda\bigg)|\bar{\mathcal{N}}\rangle,
\quad |\bar{\mathcal{N}}\rangle =\prod_{\text{odd }l}
c_{l\uparrow}^\dagger c_{l\downarrow}^\dagger |0\rangle ,
\label{E39}
\end{equation}

\noindent
is also an eigenvector of $H$ with the same energy eigenvalue
(\ref{E35}). The bosonic operators $\phi_{\text{e}}$ and
$\phi_{\text{o}}$ turn to creation operators when operating on
$|\tilde{g}\rangle$.

\begin{figure}[h!]
\begin{center}
\includegraphics[width=8.5cm]{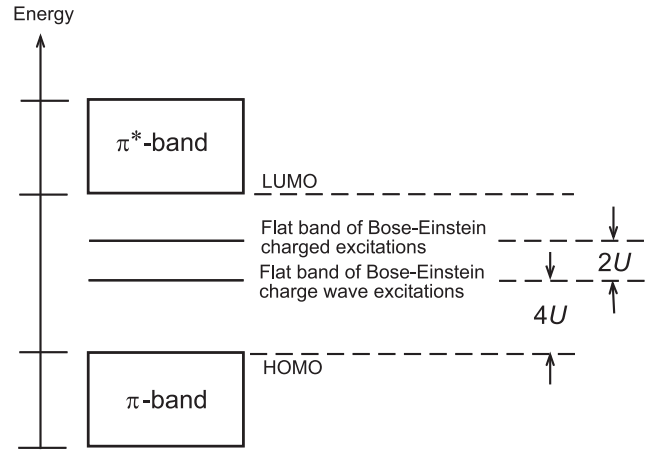}
\caption{\label{Fig1} The $\pi$--$\pi$ repulsion give rise to two
energy levels of Bose--Einstein excitations whose energies are $4U$ and
$6U$ above the energy of the highest occupied molecular orbital (HOMO).}
\end{center}
\end{figure}

Figure \ref{Fig1} shows a schematic diagram of the energy spectrum of
the molecular chain. The $\pi$--$\pi$ repulsion yields a novel energy
level of Bose--Einstein excitations located in between the energies of
the HOMO and the LUMO. Hence the conduction band in this framework is 
not constituted by electrons in extended anti--bonding states, but of
electronic pairs excited to the novel flat band of energy $4U$.
Certainly the system have other degrees of freedom, additional to the
ones described by the operators $\phi_{\text{e}}$ and
$\phi_{\text{o}}$, e.~g.~the solitons firstly invoked by Su, Schrieffer
and Heeger \cite{Su1,Su2} for polyacetilene, and then generalized to
other conjugate polymers \cite{Heeger}. They give rise to the flat band
of energy $6U$ in Figure \ref{Fig1}, which will be adressed with some
detail after showing how the excitations associated with the operators
$\phi_{\text{e}}$ and $\phi_{\text{o}}$ transport momentum and energy.
As it has been assumed that one--electron processes involve too large
activation energies, the momentum operator $\vec{P}$ must be written as
a two--particle operator, in the way

\begin{equation}
\begin{aligned}
\vec{P}=
& -\frac{i\hbar}{2}\int d^3\vec{r'}d^3\vec{r}\,\Psi^\dagger(\vec{r'},t)
\Psi^\dagger(\vec{r},t)(\nabla'\\
& +\nabla) \Psi(\vec{r},t)\Psi(\vec{r'},t)
+\,\text{adjoint operator},
\label{E40}
\end{aligned}
\end{equation}

\noindent
when written in terms of the electron field operator

\begin{equation}
\begin{aligned}
\Psi(\vec{r},t)=\sum_{ls}c_{ls} w_s(\vec{r}-la\hat{\imath}).
\label{E41}
\end{aligned}
\end{equation}

\noindent
Here $\nabla'$ and $\nabla$ are the gradient operators with respect to
$\vec{r'}$ and $\vec{r}$, and $w_s(\vec{r}-la\hat{\imath})$ represents
the one--particle wave function of an electron in a $\pi$ covalent
state at site $l$, $a$ is the distance between two adjacent sites of
the polymer chain, and $\hat{\imath}$ is the unitary vector along the
chain direction. Because of analytical reasons and the small overlap of
functions $w_s$ centered in neighboring sites 

\begin{equation}
\int d^3\vec{r}\,w_s^*(\vec{r}-l'a\hat{\imath})
\nabla w_s(\vec{r}-la\hat{\imath})=
\begin{cases}
0,\text{ if } l=l'\\
q\,\hat{\imath}\text{ if } l'=l+1\\
-q^*\,\hat{\imath}\text{ if } l'=l-1\\
0, \text{ otherwise}.
\end{cases}
\label{E42}
\end{equation}

\noindent
Inserting Eqs.~(\ref{E41}) and (\ref{E42}) as they are written into
Eq.~(\ref{E40}), the resulting expression for $\vec{P}$ finally reduces
to the same standard equation for the one--particle momentum operator.
However, one must recall that the backbone of the polymer is not rigid
and the alternating occupied and unoccupied $\pi$--orbitals should
cause a dimerization of the polymer chain. Hence the occupied and
virtual $\pi$--orbitals are expected to have a finite difference,
simply because of the broken periodicity of the distance between the
positive charges involved in the chemical bonds. In Eq.~(\ref{E42}) the
parameter $a$ takes slightly different values if the accompanying index
$l$ is even or odd. When taking this into consideration the momentum
operator splits into a one--particle term and a two--particle one,
taking the general form

\begin{equation}
\begin{aligned}
& \vec{P}=-i\hbar\,q\,\hat{\imath}
\sum_{ls}\big( c_{l+1s}^\dagger c_{ls}
-c_{ls}^\dagger c_{l+1s}\big)\\
& -i\hbar\,\gamma q\,\hat{\imath}
\sum_l\big( c_{l+1\uparrow}^\dagger c_{l+1\downarrow}^\dagger
c_{l\downarrow}c_{l\uparrow}
-c_{l\uparrow}^\dagger c_{l\downarrow}^\dagger
c_{l+1\downarrow}c_{l+1\uparrow}\big),
\label{E43}
\end{aligned}
\end{equation}

\noindent
where $\gamma$ is a coefficient proportional to the shift $\delta a$
in the bond lengths of the dimerized chain. The first term of $\vec{P}$
in general destroys pairs and the second one always conserves them. As
$H$ and the eigenstates (\ref{E24}), (\ref{E37}) and (\ref{E39})
involve just paired electrons, consistently with the principle that
one--electron processes involve too large activation energies we can
retain only the second term and write $\vec{P}$ as

\begin{equation}
\begin{aligned}
\vec{P} & =-i\hbar\,\gamma q\,\hat{\imath}
\sqrt{\frac{N}{2}}\big(\phi_{\text{e}}^\dagger
-\phi_{\text{o}}^\dagger -\phi_{\text{e}}+\phi_{\text{o}}\big).
\label{E44}
\end{aligned}
\end{equation}

In the Heisenberg picture

\begin{equation}
\frac{d^2P}{dt^2}=-\frac{1}{\hbar^2}[H,[H,P]].
\label{E45}
\end{equation}

\noindent
Replacing Eqs.~(\ref{E44}) and (\ref{E20}) one has that

\begin{equation}
\frac{d^2P}{dt^2}=-\frac{4U^2}{\hbar^2}P.
\label{E46}
\end{equation}

\noindent
Eq.~(\ref{E46}) shows that the operators $\phi_{\text{e}}^\dagger$
and $\phi_{\text{o}}^\dagger$ excite modes of the collective motion of
the charges involved in the $\pi$--bonds present in the chain. The
collective oscillation involves charge displacement and has an angular
frequency $\omega =2U/\hbar$, independent of the length of the polymer
chain. Hence resonances favouring charge transfers with the
neighbouring chains are expected to occur.

The model worked out in the preceding paragraphs shows that the
dynamical effect of the $\pi$--$\pi$ repulsion in conjugate polymers is
a set of degenerate excited states able of transporting charge along
the polymer chain. The novel states obey Bose--Einstein statistics and
have energy eigenvalues forming a flat band located at an energy $4U$
over the ground state, in the gap between the HOMO and the LUMO. The
model assumed in this step is a natural reply to the question of how
the repulsion between the $\pi$--orbitals may affect the dynamics of
the system, however is yet a bit too simple. An important conclusion
is that the single frequency $\omega =2U/\hbar$, common to all the
polymeric chains in a sample, no matter their lengths, which is
expected to produce resonances that enhance the probability of mutual
charge transfers.

\begin{figure}[h!]
\begin{center}
\includegraphics[width=8cm]{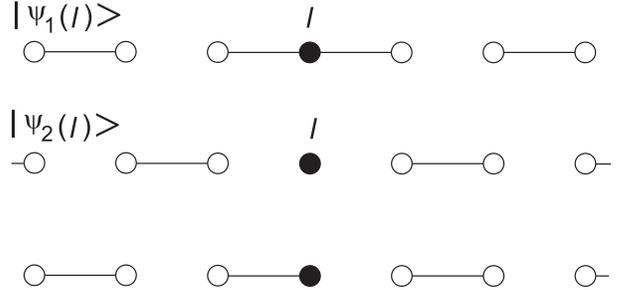}
\caption{\label{Fig2} Circles represent monomers and lines the
$\pi$--bonds. The black filling of a circle is only for reference. The
two first lines represents the states $|\psi_1\rangle$ and
$|\psi_2\rangle$, the third line shows the configuration of one of the
two degenerate ground states.}
\end{center}
\end{figure}

The metallic phase is constituted by a set of excited states similar to
those introduced by Su, Schrieffer and Heeger \cite{Su1,Su2,Heeger}. At
a site $l$ the occupation of the $\pi$--orbitals changes from even to
odd sites, or vice--versa. The monomer in between marks a frontier
between the two degenerate ground states, $|g\rangle$ and
$|\bar{g}\rangle$, prevailing at each side of it. The energy of such a
state can be inferred from an heuristic argument instead of a formal
one. Consider first a class of excited states of the form $(|\psi_1(l)
\rangle\pm |\psi_2(l)\rangle)/\sqrt{2}$, where $|\psi_1(l)\rangle$
and $|\psi_2(l)\rangle$ are the states represented schematically in
Figure \ref{Fig2}. As the interaction between the orbitals is
circumscribed to nearest neighbors, for small enough $\alpha$ the
mutual perturbation of the states in the two sides of $l$ is small.
Hence the contribution to the energy of both sides sum approximately
$4U$, and the central parts of $|\psi_1(l)\rangle$ and $|\psi_2(l)
\rangle$ contribute $4U$ and 0, in the zeroth order in $\alpha$. Hence
the mean energy of the state is $6U$. This class of excited states
configures the flat band of charged excitations represented in Figure
\ref{Fig1}. The excitations have charge $\pm 2e$ and an energy $6U$
over the energy of the highest occupied molecular orbital (HOMO). The
metallic phase is constituted by the set of linear combinations of the
form  

\begin{equation}
\begin{aligned}
&|k\rangle =\sqrt{\frac{1}{N}}\sum_{l=-N/2}^{N/2}\exp(ikl)
|\psi_{1,2}(l)\rangle ,
\quad k=\frac{2\pi}{N}n,\\
&\quad n=-\frac{N}{2},\dots,-1,0,1,2,\dots,\frac{N}{2}.
\end{aligned}
\label{E47}
\end{equation}

The predictions made here on the energy bands can be readily identified
in the light absorption spectra. The two flat bands of Figure \ref{Fig1}
are quite characteristic because their energies are in the ratio 3/2.
A recent paper \cite{Chen} shows the measured absorption spectra of
electromagnetic radiation of wavelengths in the interval 300--800 nm of
the conjugated polymer FBT-Th$_4$(1,4). The target polymer was in solid
films of 60 nm thick and solutions of chlorobenzene and dichlorobenzene
with very similar results. Two main maxima at $\lambda_1=692\text{ nm}$
and $\lambda_2=453\text{ nm}$ were observed in the film. The energies in
eV are $\epsilon_1=1.792\text{ eV}$ and $\epsilon_2=2.737\text{ eV}$ and
their ratio is

\begin{equation}
\frac{\epsilon_2}{\epsilon_1}=1.527 ,
\label{E48}
\end{equation}

\noindent
which is quite close to the predicted value $3/2$. The precision of the
agreement is quite unexpected because the real energy of the
transitions are shifted from the spectral maxima by a Stokes shift.
However the Stokes shifts are expected to be very similar for the two
maxima, which can explain such precision. The rest of the experimental
spectrum agrees very well with the scheme of Figure \ref{Fig1}, with
the maxima in between a plateau and a null region evidencing the energy
gap. 

More recent observations of the UV--Vis excitation spectra of many
conjugated polymers of interest show results similar to those of Chen
et al. \cite{Chen}, with the ratio of Eq.~(\ref{E48}) ranging from
approximately 1.4 to 1.7 \cite{Yao,Mangalore,Li,Lee,Park}. The spectra
show two main peaks centered typically at $\lambda_1\approx 700\,
\text{nm}$ and $\lambda_2\approx 450\,\text{nm}$ with full half width
of about 100 nm. The peak of higher wavelength has always a shoulder.
In general the spectra exhibit two main peaks clearly attributable to
the two flat bands derived from the model solved here. The obtention of
precise values for the discrete energy levels over the energy of the
HOMO from the UV--Vis excitation spectra demands to deconvolute the
vibrational modes, which widens the spectral maxima \cite{LagosParedes}.

\end{document}